\documentclass{article}
\usepackage{authblk,amssymb,amsmath,hyperref}

\usepackage[a4paper,left=3cm,right=3cm,top=3cm,bottom=3cm]{geometry}

\usepackage{cite}

\usepackage[T1]{fontenc}
\usepackage{lmodern}

\usepackage[frozencache]{minted}
\def\coqin#1{\mintinline{ssr}{#1}}
\setminted[ssr]{escapeinside=88}

\usepackage{tikz}

\pgfdeclarelayer{bg}    
\pgfsetlayers{bg,main}  

\usepackage{dsfont}

\def\coq{\textsc{Coq}}
\def\analysis{\textsc{MathComp-Analysis}}
\def\mathcomp{\textsc{MathComp}}
\def\hb{\textsc{Hierarchy-Builder}}

\def\FTC{FTC}
\def\lebFTC{\FTC{} for Lebesgue integration}
\def\LDT{Lebesgue Differentiation theorem}
\def\HL{Hardy-Littlewood}

\title{A Comprehensive Overview of the Lebesgue Differentiation Theorem in Coq}

\def\newterm#1{\textsl{#1}}

\def\mydef{\overset{\textrm{def}}{=}}
\def\myae{\overset{\textrm{a.e.}}{=}}

\def\indic#1{\mathds{1}_{#1}}

\def\sec#1{\S{}#1}
\def\sect#1{Sect.~#1}

\def\rhs{right-hand side}

\author[1]{Reynald Affeldt}

\author[2]{Zachary Stone}

\affil[1]{National Institute of Advanced Industrial Science and Technology (AIST), Tokyo, Japan}

\affil[2]{The MathComp-Analysis development team, Boston, MA, USA}

\begin{document}

\date{}

\maketitle

\begin{abstract}
Formalization of real analysis offers a chance to rebuild traditional
proofs of important theorems as unambiguous theories that can be
interactively explored.
This paper provides a comprehensive overview of the Lebesgue
Differentiation Theorem formalized in the Coq proof assistant,
from which the first Fundamental Theorem of Calculus (FTC) for the
Lebesgue integral is obtained as a corollary.
Proving the first FTC in this way has the advantage of decomposing
into loosely-coupled theories of moderate size and of independent
interest that lend themselves well to incremental and collaborative
development.
We explain how we formalize all the topological constructs and all the
standard lemmas needed to eventually relate the definitions of
derivability and of Lebesgue integration of MathComp-Analysis, a
formalization of analysis developed on top of the Mathematical
Components library.
In the course of this experiment, we substantially enrich
MathComp-Analysis and even devise a new proof for Urysohn's lemma.
\end{abstract}

\section{Introduction}
\label{sec:introduction}

\def\inte#1#2#3{\int_{#2} #3 (\textrm{\textbf{d}}\,#1)}

The formalization of the Fundamental Theorem of Calculus
(\FTC{}) for the Lebesgue integral in the \coq{} proof
assistant~\cite{coqrefman} is an ongoing work as part of the development of \analysis{}~\cite{analysis},
a library for analysis that extends the Mathematical
Components library~\cite{mathcompbook} with
classical axioms~\cite[\sec{5}]{cohen2018jfr}.
Besides mathematics, \analysis{} has also been used to
formalize programming
languages~\cite{affeldt2023cpp,zhou2023popl,saito2023aplas}.

The first \lebFTC{} can be stated as follows: for $f$ integrable
on~$\mathbb{R}$, $F(x)\mydef\inte{\mu}{t\in]-\infty,x]}{f(t)}$ is
differentiable and $F'(x) = f(x)$ almost-everywhere (hereafter, a.e.)
relatively to $\mu$, where $\mu$ is the Lebesgue measure.
This is different from the standard statement for the Riemann
integral, where $f$ is assumed to be continuous, making for a simple
proof.
In comparison, connecting derivation and Lebesgue integration under an
integrability hypothesis is unwieldy, even more so in \analysis{}
whose formalizations of derivation~\cite[\sec{4.5}]{cohen2018jfr} and
of the Lebesgue integral~\cite[\sec{6.4}]{affeldt2023jar} have been
unrelated so far.
They can be bridged thanks to the \LDT{} and this is appealing
for two reasons. First, it is a useful theorem in itself: the first \FTC{}
is a consequence, as well as other results such as Lebesgue's density
theorem. Second, we can decompose its proof in several results: this
provides a way to incrementally enrich the theories of
\analysis{}. We think that this approach is an instance of a more
generic way to tackle formalization of mathematics: find a path
through the literature to present many key results as easy
consequences of a central, technical lemma with rather weak
assumptions.
Incidentally, such a fine-grained decomposition also provides a
practical way to monitor formalization progress.

In terms of formalization in a proof assistant, our contributions are
as follows:
\begin{itemize}
\item We provide the first formalization of the first \lebFTC{} in \coq{}.
\item We bring to \coq{} several standard lemmas and theorems
  of measure theory: Vitali's lemmas and theorem, a theory of \HL{}'s
  operator, Urysohn’s lemma, Ergorov's, Lusin's, Tietze's theorems,
  and the \LDT{}.
  In particular, among these results, Urysohn's lemma is given an
  original proof.
  We also improve the \analysis{} support for topology (lower
  semicontinuity, normal spaces, subspaces, etc.) and for real
  functions (by extending the theory for $\limsup/\liminf$).
  The formalization of the first \lebFTC{} provides a strong evidence
  that these pieces of formalization can indeed be combined to achieve
  a large result.
\end{itemize}

Another intent with this paper is to produce an informative document
for potential users of the topology and measure theories of
\analysis{}.
The whole library is still under development in the sense that not all
notions are formalized as we would like them to be, often to cope with
temporary limitations of the available tooling.
Yet, we did observe that it is already a useful tool.  For example, we
could use it to revisit the proof of Urysohn's lemma by producing an
original proof. Also, we had to clarify a few proof steps that are
often hand-waved in lecture notes: typically, generalizations from
lemmas stated for bounded cases only. Filling such gaps is almost
business-as-usual when formalizing mathematics, but we believe that it
is important to document them to better anticipate similar gaps in the
future.
For these reasons, we think that our formalization of the \FTC{}
provides a nice milestone to document \analysis{}.

\paragraph*{Outline}
Regarding mathematical proofs, we stay at the level of a bird's-eye
view and instead focus on the main aspects of the formalization.
We start by recalling the basics of \analysis{} in
\sect{\ref{sec:preliminaries}}.
We explain the formal statement of the \LDT{}
in~\sect{\ref{sec:ldt_statement}} and provide an overview of its proof
in \sect{\ref{sec:ldt_proof}}.
To formalize this proof, we extend \analysis{} with new topological
constructs in~\sect{\ref{sec:extension_analysis}} and with basic but new
measure-theoretic lemmas in \sect{\ref{sec:more_measure_theory}}.
In the particular case of Urysohn's lemma, we explain the formalization
of an original proof in \sect{\ref{sec:urysohn}}.
The main intermediate lemmas of the \LDT{} are the purpose of
\sect{\ref{sec:density}} and \sect{\ref{sec:vitali_hardylittlewood}}.
Finally, we apply the \LDT{} to the proof of the first \lebFTC{} and
to the proof of Lebesgue's density theorem in \sect{\ref{sec:applications}}.
We review related work in \sect{\ref{sec:related_work}} and conclude in
\sect{\ref{sec:conclusions}}.

\section{Background on \analysis}
\label{sec:preliminaries}

\def\mylim#1#2#3#4{#1 \xrightarrow[ #3 \to #4 ]{} #2}

\def\msub{$\subseteq$} 
\def\minter{$\cap$} 
\def\munion{$\cup$} 
\def\mcplt{${}^\complement$} 
\def\mprod{$\times$} 
\def\mdiff{$\setminus$} 

\analysis{} is built on top of \mathcomp{} and reuses several of its
theories. We use in particular the following notations, which are
traditionally in ASCII in \mathcomp{} libraries.  The successor function of natural numbers is
noted \coqin{.+1}, multiplication of a natural number by 2 is noted
\coqin{.*2}.
A multiple conjunction is noted \coqin{[/\ P1, P2, ... & Pn]}.
Function composition is noted \coqin{\o}.
Point-wise equality between two functions is noted \coqin{=1}.
Point-wise multiplication of two functions $\texttt{f}$ and $\texttt{g}$ is noted \coqin{f \* g}.
The notation \coqin{f ^~ y} is for the function $\lambda x. \texttt{f}\,x\,\texttt{y}$.
Intervals are noted \coqin{`]a, b[}, \coqin{`]a, b]}, etc.
%
%
One can inject a natural number \coqin{n} into a ring with the notation \mintinline{ssr}{n
The inverse of a field element \coqin{r} is noted \coqin{r^-1}.
The norm of \coqin{x} is noted \coqin{`|x|}.
The inclusion between two lists \coqin{s} and \coqin{r} is noted
\coqin{{subset s <= r}}.
We write \coqin{x \in s} when an element~\coqin{x} belongs to the list~\coqin{s}.

\analysis{} comes with library support for set theory.
Given a type \coqin{T}, \coqin{set T} is the type of sets of elements of type \coqin{T}.
The notation \coqin{[set: T]} is for the set of all the elements of
type \coqin{T};
the singleton set containing \coqin{x} is noted \coqin{[set x]};
and \coqin{set0} represents the empty set.
To unambiguously improve readability, we use standard \LaTeX{} notations 
instead of ASCII for the set-theoretic operations
set inclusion (\msub), 
set intersection (\minter), 
set union (\munion), 
set difference (\mdiff), and 
the product of sets (\mprod). 
The complement of a set \coqin{A} is noted \coqin{A}\mcplt. 
(If necessary, see \cite[Table~2]{affeldt2023jar} for a list-up of the
corresponding ASCII notations.)
Set difference with a singleton is noted \coqin{A `\ x}:
it is a shortcut for \mintinline{ssr}{A 8\mdiff8 [set x]}.
The image by a function~\coqin{f} of a set \coqin{A} is written \coqin{f @` A}.
The set $\{f(x)\,|\,x\in A\}$ defined by comprehension is noted \coqin{[set f x | x in A]}.
A list \coqin{s} can be turned into a set with the notation \coqin{[set` s]}.
The notation \coqin{A !=set0} means that the set \coqin{A} is not empty.
A family \coqin{F} of pairwise-disjoint sets indexed by \coqin{D} is noted
\coqin{trivIset D F}.
The characteristic function over a set \coqin{A} is noted~\coqin{\1_A}.

\analysis{} extends the numeric types of \mathcomp{} with the type \coqin{realType} for reals.
When \coqin{R} is a numeric type, \coqin{\bar R} is the numeric type extended with \coqin{-oo} and~\coqin{+oo},
so that when \coqin{R : realType}, \coqin{\bar R} corresponds to $\overline{\mathbb{R}}$.
One can inject a numeric value \coqin{r} into the corresponding type of extended numbers
with the notation \mintinline{ssr}{r
An extended number \coqin{x} can be projected to the corresponding numeric value
by \coqin{fine x} (which is \coqin{0} when \coqin{x} is $\pm\infty$).
The supremum of a set of extended reals \coqin{A} is noted \coqin{ereal_sup A}.
In this paper, the variable \coqin{R} has type \coqin{realType} unless
stated otherwise.

Like several other
libraries~\cite{holzl2013itp,boldo2015mcs,mathlib2020cpp}, \analysis{}
uses filters to formalize topology.
For example, we note \coqin{\oo} the filter consisting of the set of
predicates over natural numbers that are eventually true;
\coqin{x^'} the \newterm{deleted neighborhood filter} of \coqin{x},
i.e., the set of neighborhoods of \coqin{x} from which \coqin{x} is
excluded;
\mintinline{ssr}{x^'+} for \newterm{right filters}, i.e., the filters
of neighborhoods of~\coqin{x} intersected with $]\texttt{x},+\infty[$,
and similarly for the \newterm{left filters} note
\mintinline{ssr}{x^'-}.
Filters are associated to elements of a given type upon the definition of a \newterm{filtered type}.
The convergence statement $\mylim{f(x)}{\ell}{x}{a}$ is noted
\coqin{f x @[x --> a] --> l}; the limit of a filter \coqin{F} is noted
\coqin{lim F} \cite[\sec{2.3}]{cohen2018jfr}.
Topological spaces are built on top of filtered types and their type
is \coqin{topologicalType}. In a topological space, the set of
neighborhoods of \coqin{x} is \coqin{nbhs x}.
Mathematical structures such as topological spaces are defined and
instantiated using a \coq{} extension called
\hb{}~\cite{cohen2020fscd}. With this tool, interfaces are defined as
so-called \newterm{factories} whose definition generates constructors
to build instances.  See \cite[\sec{3.1}]{affeldt2023jar} for a quick
reference to \hb.
The predicates \coqin{open}, \coqin{closed}, \coqin{closure}, and
\coqin{compact} correspond to the eponymous topological notions.
The type of uniform spaces is \coqin{uniformType}.
Given a uniform space \coqin{M},
entourages are objects of type \coqin{set (set (M * M))}
that satisfy the axioms of uniform space for \coqin{M}.
One of the axiom of uniform spaces guarantees that for each entourage~\coqin{E},
there is an entourage \coqin{V} with
$\{(x,z)\,|\,\exists y,\,(x,y) \in V \,\land\, (y,z) \in V\} \subseteq E$.
We denote this entourage with \coqin{split_ent(E)}.
\analysis{} also provides pseudometric spaces in which a ball is noted
\coqin{ball}; over the real line, a ball is a centered open interval.
The type of sequences (indexed by natural numbers) over \coqin{T} is noted~\coqin{T^nat}.

The basics of measure theory in \analysis{} is documented in previous work~\cite{affeldt2023jar}.
A \coqin{measurableType} is a type equipped with a structure of $\sigma$-algebra.
Given a measurable type~\coqin{T} and \coqin{A} of type \coqin{set T},
we write \coqin{measurable A} when the set \coqin{A} is measurable.
The fact that the extended real-valued function $f$ is measurable over
$D$ is noted \coqin{measurable_fun D f}.
For a measure~\coqin{mu}, the fact that \coqin{f} is integrable over \coqin{D} is noted \coqin{mu.-integrable D f}.
The integral $\inte{\mu}{x\in A}{f(x)}$ is noted \mintinline{ssr}{\int[mu]_(x in A) f x}.
When \coqin{A} is negligible for a measure~\coqin{mu}, we write \coqin{mu.-negligible A}.
The fact that the predicate \coqin{P} holds a.e.\ relatively to
\coqin{mu} is noted \coqin{{ae mu, forall x, P x}}.

\def\measurable#1{\texttt{measurable}(#1)}

\section{Statement of the \LDT}
\label{sec:ldt_statement}

\def\ball#1#2{\textrm{B}_{#1}(#2)}

\def\iavg#1#2{\frac{1}{\mu(#2)}\inte{\mu}{y\in #2}{|#1(y)|}}
\def\iavgnotation#1#2{\left[{#1}\right]_{#2}}
\def\davg#1#2#3{\iavg{\lambda y.\ #1(y) - #1(#2)}{\ball{#3}{#2}}}
\def\davgnotation#1#2#3{\overline{{#1}_{\ball{#3}{#2}}}}

Let us note $\iavgnotation{f}{A} \mydef \iavg{f}{A}$ the average of a
real-valued function $f$ over the set~$A$.  Using the existing
notations of \analysis, we formalize this notion as follows:
\begin{minted}{ssr}
Definition iavg (f : R -> R) (A : set R) :=
  (fine (mu A))^-1%:E * \int[mu]_(y in A) `| (f y)%:E |.
\end{minted}
Recall from \sect{\ref{sec:preliminaries}} that
\mintinline{ssr}{
and that \coqin{fine} performs a corresponding projection.
We also introduce the notation
$\davgnotation{f}{x}{r} \mydef \iavgnotation{\lambda y.f(y)-f(x)}{\ball{r}{x}}$
where $\ball{r}{x}$ is a ball centered at $x$ of radius~$r$:
\begin{minted}{ssr}
Definition davg (f : R -> R) (x r : R) := iavg (center (f x) \o f) (ball x r).
\end{minted}
The \coqin{center c} function is $\lambda y.y-c$.

Given a real-valued function $f$, a \newterm{Lebesgue point\/} is a
real number~$x$ s.t.\
$\mylim{\davgnotation{f}{x}{r}}{0}{r}{0^+}$.
Reusing the Lebesgue measure (hereafter~\coqin{mu}) formalized
in previous work~\cite[\sec{5.2}]{affeldt2023jar}, we formally define Lebesgue points:
\begin{minted}{ssr}
Definition lebesgue_pt (f : R -> R) (x : R) :=
  davg f x r @[r --> 0^'+] --> 0.
\end{minted}
%
%
Note the use of right filters (\sect{\ref{sec:preliminaries}}) to define the
fact that \coqin{r} tends to~$0^+$.
The \LDT{} states that, for a real-valued, locally-integrable function
$f$ (i.e., integrable on compact subsets of its domain), we have
Lebesgue points a.e.:
\begin{minted}{ssr}
Lemma lebesgue_differentiation (f : R -> R) : locally_integrable [set: R] f
  -> {ae mu, forall x, lebesgue_pt f x}.
\end{minted}
Being locally integrable can be defined as the following conjunction:
\begin{minted}{ssr}
Definition locally_integrable (D : set R) (f : R -> R) :=
  [/\ measurable_fun D f, open D & forall K, K 8\msub8 D -> compact K ->
        \int[lebesgue_measure]_(x in K) `|f x|%:E < +oo].
\end{minted}
In fact, this definition specializes in a locally compact space such
as $\mathbb{R}$ to the conjunction of \coqin{open D} and
\begin{minted}{ssr}
forall x, D x -> exists U, open_nbhs x U /\ mu.-integrable U (EFin \o f)
\end{minted}
where \coqin{open_nbhs} is a predicate for open neighborhoods; yet, we
stuck to the previous definition from our reference textbook~\cite{li2022}.

\section{Proof of the \LDT{}}
\label{sec:ldt_proof}

\def\HLmax#1{{\textrm{HL}(#1)}}

The first step of the proof of the \LDT{} is to reduce the problem
to functions $f_k \mydef f\indic{B_k}$ with $B_k\mydef \ball{2(k+1)}{0}$:
\begin{minted}{ssr}
Lemma lebesgue_differentiation_bounded (f : R -> R) :
  let B k := ball 0 k.+1.*2%:R in let f_ k := f \* \1_(B k) in
  (forall k, mu.-integrable [set: R] (EFin \o f_ k)) ->
  forall k, {ae mu, forall x, B k x -> lebesgue_pt (f_ k) x}.
\end{minted}
This problem reduction is often hand-waved in lecture notes
(Schwartz's presentation is one exception~\cite[eqn (5.12.101)]{schwartz1997}).

Second, instead of proving for all $k$ that we have a.e.\ Lebesgue
points over $B_k$, it is sufficient to prove that the set
$A_k(a) \mydef B_k \cap \left\{ x\,|\, a < \limsup_{r\to 0}
  \davgnotation{f_k}{x}{r} \right\}$ is negligible for all $a>0$,
i.e.:
\begin{minted}{ssr}
(* local context omitted *)
============================
mu.-negligible (B k 8\minter8 [set x | a%:E < (f_ k)^* x])
\end{minted}
where $h$\coqin{^*}\;\coqin{x} is a (local) notation for $\limsup_{r\to 0}\davgnotation{h}{\mathtt{x}}{r}$.

For this last step, the idea is to exhibit a sequence of continuous
functions $g_i$ such that
$$A_k(a)
\subseteq
\bigcap_n B_k \cap \bigl(
  \underbrace{\left\{x \,|\, f_k(x) - g_n(x) \geq a/2\right\}}_{(a)}
  \cup
  \underbrace{\left\{x \,|\, \HLmax{f_k(x) - g_n(x)} > a/2\right\}}_{(b)}
\bigr)
$$
where
$\HLmax{f}(x)\mydef \sup_{r>0}\{ \iavgnotation{f}{\ball{r}{x}} \}$ is the \HL{} operator.
We can show that the measure of the \rhs{} is null.
To deal with $(a)$, we use Markov's inequality (i.e., the fact that
$\mu(\{x\,|\;|f(x)|\geq a\})\leq \frac{1}{a}||f||_1$ for $a>0$ and a
measure $\mu$, where $||\cdot||_1$ is the $L^1$ norm) and the fact
that continuous functions are dense in $L^1$, whose proof requires several
standard results of measure theory (in particular, Urysohn's lemma).
To deal with $(b)$, we need the \HL{} maximal inequality, which in
turns relies on Vitali's covering lemma.

Figure~\ref{fig:ldt_graph} shows the main lemmas that we add to
\analysis{} to formalize this proof.

\def\bounded{{\color{blue}\small(bounded)}}

\begin{figure}[t]
\centering
\begin{tikzpicture}
\begin{scope}[every node/.style={draw,fill=white,rounded corners=0.3em},
              every path/.style={-stealth},
              boundedgen/.style={dashed,blue}]
  \node (ftc) {\lebFTC{} (\sec{\ref{sec:ftc}})};
  \node (density) [right of=ftc,xshift=14em]  {Lebesgue's density theorem (\sec{\ref{sec:lebesgue_density}})};
  \node (lebdiffthm) [above of=ftc,xshift=7em,thick] {\LDT{} (\sec{\ref{sec:ldt_statement}})};
  \node (lebdiffthmrestr) [above of=lebdiffthm] {\LDT{} \bounded{} (\sec{\ref{sec:ldt_statement}})};
  \node (hardylittle) [above left of=lebdiffthmrestr,xshift=-4.5em,yshift=0.8em]
    {\begin{tabular}{c}
     \HL{} maximal inequality (\sec{\ref{sec:hardylittlewood}})
     \end{tabular}};
  \node (vitalilemma) [above of=hardylittle,xshift=-2.3em] {Vitali's lemma (\sec{\ref{sec:vitali}})};
  \node (dense) [above right of=lebdiffthmrestr,xshift=12.8em,yshift=0.8em]
    {\begin{tabular}{c}
     Continuous functions are dense in $L^1$ (\sec{\ref{thm:dense}})
     \end{tabular}};
  \node (lusin) [above right of=dense,xshift=-6.3em,yshift=0.9em] {Lusin's thm (\sec{\ref{thm:lusin}})};
  \node (innerregularity) [above left of=lusin,xshift=-7em,yshift=1em]
    {\begin{tabular}{c}Inner regularity (\sec{\ref{sec:inner_regularity}}) \end{tabular}};
  \node (innerregularitybounded) [above of=innerregularity,xshift=3em,yshift=0.1em]
    {\begin{tabular}{c}Inner regularity \bounded{} (\sec{\ref{sec:inner_regularity}}) \end{tabular}};
  \node (outerregularity) [above of=innerregularitybounded] {Outer regularity (\sec{\ref{sec:outer_regularity}})};
  \node (egorov) [above right of=lusin,xshift=0.5em,yshift=0.9em] {\begin{tabular}{c}Egorov's\\thm (\sec{\ref{sec:egorov}})\end{tabular}};
  \node (tietze) [above left of=dense,xshift=7.6em,yshift=0.7em] {Tietze's thm (\sec{\ref{thm:tietze}})};
  \node (urysohn) [above of=tietze,xshift=-0.1em,yshift=0.4em] {\begin{tabular}{c}Urysohn's\\lemma (\sec{\ref{sec:urysohn}})\end{tabular}};

  \draw (vitalilemma) -- (hardylittle);
  \draw (hardylittle) -- (lebdiffthmrestr);
  \draw (dense) -- (lebdiffthmrestr);
  \draw (lusin) -- (dense);
  \begin{pgfonlayer}{bg}
    \draw (innerregularitybounded) -- (lusin);
    \draw (innerregularity) -- (hardylittle);
  \end{pgfonlayer}
  \draw (outerregularity) -- (innerregularitybounded);
  \draw[boundedgen] (innerregularitybounded) -- (innerregularity);
  \draw (egorov) -- (lusin);
  \draw (tietze) -- (dense);
  \draw (urysohn) -- (tietze);
  \draw (lebdiffthm) -- (ftc);
  \draw[boundedgen] (lebdiffthmrestr) -- (lebdiffthm);
  \draw (lebdiffthm) -- (density);
\end{scope}
\end{tikzpicture}
\caption{Overview of the proof of the \LDT{} and derived results\\
\footnotesize (dashed arrows indicate generalizations from
statements about bounded sets)
}
\label{fig:ldt_graph}
\end{figure}
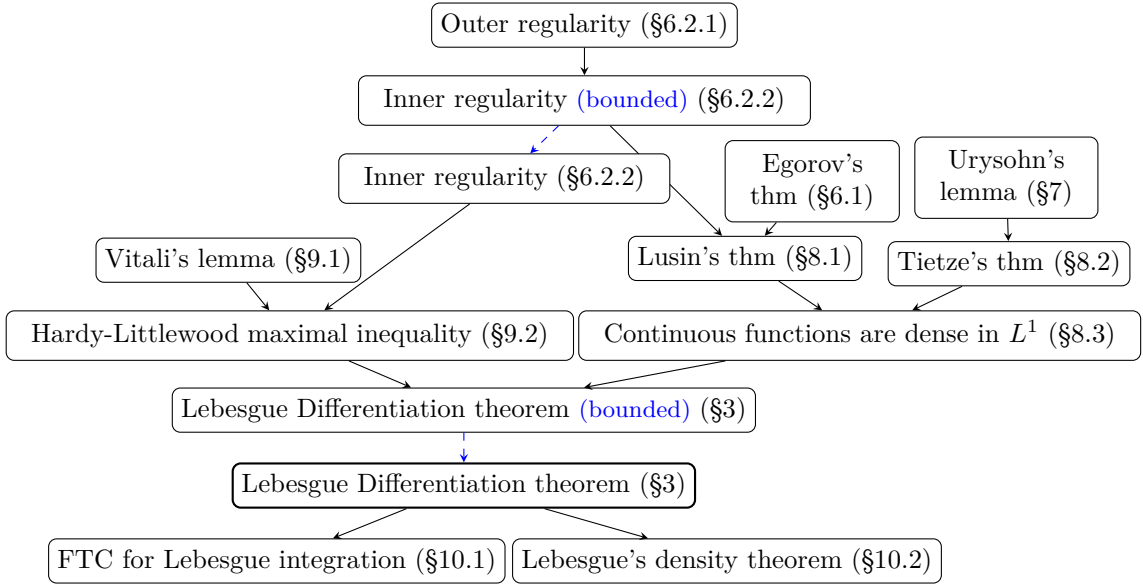

\section{Topological constructs added to \analysis{}}
\label{sec:extension_analysis}

Before explaining the formalization of the main lemmas to prove the
\LDT{}, we explain the preparatory work needed to extend the
formalization of topology of \analysis{}.
In particular, as a preliminary step to be able to state Egorov's
theorem, Lusin's theorem, and Tietze's theorem, we extend \analysis{}
with the subspace topology and with the topology of uniform
convergence.

\subsection{Subspace topologies}
\label{sec:subspace}

Given a type $T$ equipped with a topology ${\cal T}$ and a set $A$ of elements of
$T$, $\{A\cap U \;|\; U\in {\cal T}\}$ forms a subspace topology.
We formalize subspace topologies on the basis of the existing
formalization of topological space of \analysis{} and using \hb{} (see
\sect{\ref{sec:preliminaries}}).
First, we introduce an identifier \coqin{subspace} that serves as an
alias for a type \coqin{T} and which is parameterized by a set
\coqin{A}:
\begin{minted}{ssr}
Definition subspace {T : Type} (A : set T) := T.
\end{minted}
When the underlying type \coqin{T} is equipped with a topology (i.e.,
it has type \coqin{topologicalType}), we can equip the identifier
\coqin{subspace} with a structure of topology relative to
\coqin{A}. For that purpose, we start by defining a filter for a
point \coqin{x} in the subspace topology as:
(1) the restriction of the neighborhoods of \coqin{x} to \coqin{A}
(below: \coqin{within A (nbhs x)}) when \coqin{x} belongs
to~\coqin{A}, or, otherwise,
(2) the filter of the sets containing the singleton $\{ \texttt{x} \}$
(below: \coqin{globally [set x]}):
\begin{minted}{ssr}
Definition nbhs_subspace (x : subspace A) : set_system (subspace A) :=
  if x \in A then within A (nbhs x) else globally [set x].
\end{minted}
A \coqin{set_system} is simply synonymous for a set of sets.
Filters defined in this way give rise to a filtered
type~(\sect{\ref{sec:preliminaries}}). To attach the structure of
filtered type with the identifier \coqin{subspace}, it suffices to
summon \hb{} with the appropriate \newterm{factory}:
\begin{minted}{ssr}
HB.factory Record hasNbhs T := { nbhs : T -> set_system T }.
\end{minted}
A factory is represented by an interface (here, \coqin{hasNbhs}) and its definition gives rise to 
a constructor (here, \coqin{hasNbhs.Build}) that can be used to build instances, e.g.:
\begin{minted}{ssr}
HB.instance Definition _ := hasNbhs.Build (subspace A) nbhs_subspace.
\end{minted}

The filtered type associated with \coqin{subspace} can furthermore be
equipped with a topology using this other factory provided by
\analysis{}:
\begin{minted}{ssr}
HB.factory Record Nbhs_isNbhsTopological T of Nbhs T := {
  nbhs_filter : forall p : T, ProperFilter (nbhs p);
  nbhs_singleton : forall (p : T) (A : set T), nbhs p A -> A p;
  nbhs_nbhs : forall (p : T) (A : set T), nbhs p A -> nbhs p (nbhs ^~ A) }.  
\end{minted}
Indeed,
(1) one can show that \coqin{nbhs_subspace}'s are \newterm{proper
  filters} (i.e., no set in the filter is empty)
\cite[\sec{3.2.2}]{cohen2018jfr},
(2) points are contained into their neighborhoods, and
(3) given a neighborhood~$A$, the set of points of which $A$ is a neighborhood
is also a neighborhood.
It suffices to build an instance of topological type by passing to the
factory above the proofs of the facts (1), (2), and (3) (omitted here
but that can be found in the formal
development~\cite[file \coqin{topology.v}]{analysis}):
\begin{minted}{ssr}
HB.instance Definition _ := Nbhs_isNbhsTopological.Build (subspace A)
  nbhs_subspace_filter nbhs_subspace_singleton nbhs_subspace_nbhs.
\end{minted}
This is the subspace topology relative to~\coqin{A}.  Note that this
topology is discrete outside of~\coqin{A}.

In particular, we use the topology of subspace to define continuity as
follows. The notation \coqin{{within A, continuous f}} denotes
continuity of \coqin{f : subspace A -> Y}:
\begin{minted}{ssr}
Notation "{ 'within' A , 'continuous' f }" :=
  (continuous (f : subspace A -> _)).
\end{minted}
The notation \coqin{continuous} comes from
\analysis{}~\cite[file \coqin{topology.v}]{analysis} and predates this
work.
The purpose of the notation \coqin{{within A, continuous f}} is to
ensure that the continuity of \coqin{f} depends only on its values in~\coqin{A}
while \coqin{f(x + eps)} type-checks.
If \coqin{x} and \coqin{eps} have type \coqin{subspace A}, then
\coqin{f(x + eps)} is indeed well-defined.
One might naively attempt to use the sigma type \coqin{{x | A x}} with
the weak topology (i.e., the topology defined by the preimages of opens
for a given function) induced by the function \coqin{projT1 : {x : T | A x} -> T}
to define the subspace topology.
However, there is no clear way to define addition for \coqin{{x | A x}}
so that \coqin{f(x + eps)} is well-typed.

\subsection{Uniform convergence}
\label{sec:uniform_cvg}

The methodology to formalize the topology of uniform convergence is
similar to the one used to formalize subspaces.

First, let us recall how total functions are equipped with the
structure of uniform space in \analysis. There is an identifier
\coqin{arrow_uniform_type} which is an alias for the type
\coqin{U -> V} with \coqin{U : choiceType} and \coqin{V : uniformType}.
The corresponding uniform space is generated by the entourages
$\{(f,g) \,|\, \forall x : U, E (f(x), g(x))\}$ where $E$ is an
entourage of $V$ (see \coqin{fct_ent} in \cite[file \coqin{function_spaces.v}]{analysis}).
As a consequence of the appropriate \hb{} instantiation,
\coqin{arrow_uniform_type} is given the type \coqin{uniformType}.

Now, we formalize a topology whose elements are functions from the set
\coqin{A} to the type \coqin{V}. First, we introduce an identifier:
\begin{minted}{ssr}
Definition uniform_fun {U : Type} (A : set U) (V : Type) : Type := U -> V.
\end{minted}
We also introduce a notation \coqin{{uniform` A -> V}} which stands
for \coqin{@uniform_fun _ A V}.
We then introduce the (high-order) function \coqin{sigL_arrow} that
turns a function $U\to V$ between two types into a function $A\to V$
from a set to a type:
\begin{minted}{ssr}
Definition sigL_arrow {U : choiceType} (A : set U) (V : uniformType) :
  (U -> V) -> arrow_uniform_type A V := @sigL _ V A.
\end{minted}
The function \coqin{sigL} comes from \cite[file \coqin{functions.v}]{analysis}.
The identifier \coqin{uniform_fun} is then equipped with the weak
topology (\coqin{weak_topology} in \analysis) induced by
\coqin{sigL_arrow} and eventually the desired topology is simply obtained by
\newterm{copying} the structure obtained by weak topology:
\begin{minted}{ssr}
HB.instance Definition _ (U : choiceType) (A : set U) (V : uniformType) :=
  Uniform.copy {uniform` A -> V} (weak_topology (@sigL_arrow _ A V)).
\end{minted}
Copying a structure is a feature provided by \hb.  Note that we can
copy the structure using \coqin{Uniform.copy} instead of
\coqin{Topological.copy} because weak topology always inherits a
uniform structure, see the section \coqin{weak_uniform} in
\cite[file \coqin{topology.v}]{analysis}.

Then, the uniform convergence of a sequence of functions \coqin{F}
towards \coqin{f} over a set \coqin{A} can be defined.  It is the
filter inclusion of the neighborhoods of \coqin{f} in \coqin{{uniform`
    A -> V}} into the filter \coqin{F},
and this inclusion is written \coqin{cvg_to F (nbhs (f : {uniform` A -> _}))}
in \analysis.
The notation \coqin{{uniform A, F --> f}} is for the latter inclusion.

\subsection{More support for extended real-valued functions}

Besides the topological structures we explained in the previous
sections, \sect{\ref{sec:ldt_proof}} also highlights the need
to generalize \analysis{}'s theory of $\limsup/\liminf$.
Before our experiment, this theory was limited to sequences (indexed
by natural numbers) that were actually introduced to develop the
monotone convergence theorem and its
consequences~\cite[\sec{6.5}]{affeldt2023jar}.
Handling extended real-valued functions over the real numbers requires
the formalization of the following definition:
$\displaystyle \limsup_{x\to a}f(x)\mydef \lim_{\varepsilon\to 0^+}
\sup\{f(x)\,|\,x\in \ball{\varepsilon}{a}\setminus \{a\}\}$.
It can be couched in formal terms by first defining the limit superior
of a function \coqin{f} at filter \coqin{F}:
\begin{minted}{ssr}
Variables (T : choiceType) (X : filteredType T) (R : realFieldType).
Implicit Types (f : X -> \bar R) (F : set (set X)).
Definition limf_esup f F := ereal_inf [set ereal_sup (f @` V) | V in F].
\end{minted}
We can then specialize this definition to define the limit superior of
a function over the type of real numbers by using deleted neighborhood filters:
\begin{minted}{ssr}
Variable R : realType.
Implicit Types (f : R -> \bar R) (a : R).
Definition lime_sup f a : \bar R := limf_esup f a^'.
\end{minted}
This generic definition of \coqin{lime_sup} can be shown to be equivalent
to $\lim_{\varepsilon\to 0^+} \sup\{f(x)\,|\,x\in \ball{\varepsilon}{a}\setminus \{a\}\}$:
\begin{minted}{ssr}
Let sup_ball f a r := ereal_sup [set f x | x in ball a r `\ a].
Lemma lime_sup_lim f a : lime_sup f a = lim (sup_ball f a e @[e --> 0^'+]).
\end{minted}
In the course of formalizing the \LDT{}, developing the theory of
$\limsup$/$\liminf$ turned out to be a non-trivial intermission, which
revealed some quirks (now fixed) in the automatic handling of right filters using
the \coqin{near} tactics~\cite[\sec{3.2}]{cohen2018jfr} in \analysis.

\section{Egorov's theorem and regularity}
\label{sec:more_measure_theory}

The top part of Fig.~\ref{fig:ldt_graph} reveals the first
measure-theoretic results needed to prove the \LDT{}.
Lusin's theorem requires Egorov's theorem as well as the inner
regularity of the Lebesgue measure, which is also used to prove the
\HL{} maximal inequality.
In the following, \coqin{mu} is the Lebesgue measure.
We give little detail about the proofs in this section because proof
scripts are essentially textbook, they can be found in
\analysis~\cite[file \coqin{lebesgue_measure.v}]{analysis}.

\subsection{Egorov's theorem}
\label{sec:egorov}

Egorov's theorem relates convergence a.e.\ and uniform convergence
(\sect{\ref{sec:uniform_cvg}}).
Let $A$ be a bounded measurable set,
$f_k$ be a sequence of functions measurable over $A$, and
$g$ be a function measurable over $A$.
Suppose that $f_k$ convergences a.e.\ relatively to the Lebesgue
measure $\mu$ towards $g$.
Then for any $\varepsilon>0$, there exists a measurable set $B$ such
that $\mu(B)<\varepsilon$ and $f_k$ converges uniformly towards $g$
over $A\setminus B$.
Formally, assuming \coqin{T} is a measurable type:
\begin{minted}{ssr}
Lemma ae_pointwise_almost_uniform (f_ : (T -> R)^nat) (g : T -> R) A eps :
  (forall k, measurable_fun A (f_ k)) -> measurable_fun A g ->
  measurable A -> mu A < +oo ->
  {ae mu, forall x, A x -> f_ ^~ x @ \oo --> g x} ->
  (0 < eps)%R -> exists B, [/\ measurable B, mu B < eps%:E &
    {uniform A 8\mdiff8 B, f_ @ \oo --> g}].
\end{minted}
The notation for uniform convergence has been explained in \sect{\ref{sec:uniform_cvg}}.

\subsection{Regularity}
\label{sec:regularity}

Proving the outer regularity of the Lebesgue measure is a preliminary
step before proving its inner regularity.

\subsubsection{Outer regularity}
\label{sec:outer_regularity}

The Lebesgue measure is \newterm{outer regular}, which means that
it can be approximated from above by open subsets. More formally,
for every bounded measurable set $D$ and $\varepsilon>0$,
there exists an open $U \supseteq D$ such that $\mu(U\setminus D)<\varepsilon$:
\begin{minted}{ssr}
Lemma lebesgue_regularity_outer D eps :
  measurable D -> mu D < +oo -> (0 < eps)%R ->
  exists U : set R, [/\ open U , D 8\msub8 U & mu (U 8\mdiff8 D) < eps%:E].
\end{minted}
The proof is based on the definition of the Lebesgue measure as the
infimum of the measures of covers, i.e.,
$\mu(X)=\inf_F\left\{ \sum_{k=0}^\infty \mu(F_k) \,\text{\textbar}\, (\forall
  k, \measurable{F_k} ) \land X \subseteq \bigcup_k F_k\right\}$.

\subsubsection{Inner regularity}
\label{sec:inner_regularity}

Intuitively, a measure is \newterm{inner regular} when it can be
approximated from within by a compact subset.
The inner regularity of the Lebesgue measure states
that for every (bounded) measurable set $D$ and
$\varepsilon>0$, there exists a compact set
$V\subseteq D$ such that $\mu(D\setminus V) < \varepsilon$:
\begin{minted}{ssr}
Lemma lebesgue_regularity_inner D eps :
  measurable D -> mu D < +oo -> (0 < eps)%R ->
  exists V : set R, [/\ compact V , V 8\msub8 D & mu (D 8\mdiff8 V) < eps%:E].  
\end{minted}

Textbooks also resort to an alternative statement of inner
regularity. Precisely, the above statement about a bounded measurable
set can be generalized using the $\sigma$-finiteness of the Lebesgue
measure by saying that the measure of a measurable set can be
expressed as the supremum of the measure of the compact sets included
inside, i.e.,
$\mu(D) = \sup \{\mu(K) \,|\, \text{compact}(K) \land K\subseteq D\}$:
\begin{minted}{ssr}
Lemma lebesgue_regularity_inner_sup D : measurable D ->
  mu D = ereal_sup [set mu K | K in [set K | compact K /\ K 8\msub8 D]].
\end{minted}
As a matter of fact, we do use both forms in our development (Fig.~\ref{fig:ldt_graph}).

\section{A new proof of Urysohn's lemma}
\label{sec:urysohn}

The classical version of Urysohn's lemma states that a topological
space $T$ is \newterm{normal\/} (i.e., two disjoint closed sets
have disjoint open neighborhoods) if and only if, for any closed,
disjoint, non-empty subsets $A$ and $B$, there is a continuous
function $f : T \rightarrow \mathbb{R}$ such that $f(A) = \{0\}$ and
$f(B) = \{1\}$.
This result is important because it connects a purely
topological property (normality) with a purely analytic property (a
function into the reals).
Traditionally the proof involves an induction over the rationals to
explicitly construct such a function.
We do not follow the traditional proof for a technique that, we
believe, is more appealing from the viewpoint of formal verification.
Our proof has the same pattern as proving the \LDT{} first, and then
the \FTC{}:  we found that proving the right intermediate lemmas made
several results, including Urysohn's much easier.

\subsection{Urysohn's lemma using uniform separator}
\label{sec:uniformseparator}

We start by stating one of our intermediate lemmas. We first introduce
a new definition. For a topological space \coqin{T}, we define a
\newterm{uniform separator} as follows:
\begin{minted}{ssr}
Definition uniform_separator (A B : set T) := 
  exists (uT : @Uniform.axioms_ T^o) (E : set (T * T)),
    let UT := Uniform.Pack uT in [/\ 
      @entourage UT E, 
      (A 8\mprod8 B) 8\minter8 E = set0 & 
      (forall x, @nbhs UT UT x 8\msub8 @nbhs T T x)].  
\end{minted}
This says that there is a uniform structure~\coqin{UT} on~\coqin{T}
which separates~\coqin{A} and~\coqin{B}, and is coarser than the
\coqin{T} topology. This is subtly different than assuming that \coqin{T}
is a uniform space, which would imply
\coqin{forall x, @nbhs UT UT x = @nbhs T T x}, which is too strong for our purposes.
Also, \mintinline{ssr}{(A 8\mprod8 B) 8\minter8 E = set0} has a nice visual:
since an entourage is a region around the diagonal of \coqin{T * T}, 
\mintinline{ssr}{(A 8\mprod8 B) 8\minter8 E = set0} means that the region
\mintinline{ssr}{A 8\mprod8 B} is far from the diagonal.

The key result about uniform separators is:
\begin{minted}{ssr} 
Lemma uniform_separatorP {T : topologicalType} {R : realType} (A B : set T) : 
  uniform_separator A B <-> exists f : T -> R, [/\
    continuous f, 
    range f 8\msub8 `[0, 1], 
    f @` A 8\msub8 [set 0] & 
    f @` B 8\msub8 [set 1]].
\end{minted}
For the sake of readability, we defer the explanation of the proof to
the next section (\sect{\ref{sec:uniformsep}}).

The lemma just above is nearly Urysohn's lemma, but does not assume
that \coqin{T} is a normal space.  In fact, Urysohn's lemma follows
immediately from the following lemma \cite[file
\coqin{normedtype.v}]{analysis}:
\begin{minted}{ssr} 
Lemma normal_uniform_separator {T : topologicalType} (A : set T) (B : set T) : 
  normal_space T -> closed A -> closed B -> A 8\minter8 B = set0 -> 
  uniform_separator A B.
\end{minted}

The advantage of the general lemma \coqin{uniform_separatorP} is that,
besides Urysohn's lemma, we can derive other results. A
\newterm{completely regular space} $T$ is a topological space where
for every point~$x$ and closed set~$A$ with $x \notin A$ there is a
continuous function $f : T \to \mathbb{R}$ with $f(x) = 0$ and
$f (A) = \{1\}$. The classical result is that uniform spaces are
completely regular (lemma \coqin{uniform_completely_regular} in
\cite[file \coqin{normedtype.v}]{analysis}).  This follows from the
lemma \coqin{uniform_separatorP} and the following lemma:
\begin{minted}{ssr} 
Lemma point_uniform_separator {uniformType T} (x : T) (B : set T) : 
  closed B -> ~ B x -> uniform_separator [set x] B.  
\end{minted}

So, just like this paper formalizes the \LDT{} and proves the \FTC{}
among others, we find out that the lemma \coqin{uniform_sepratorP}
proves Urysohn's lemma and more.

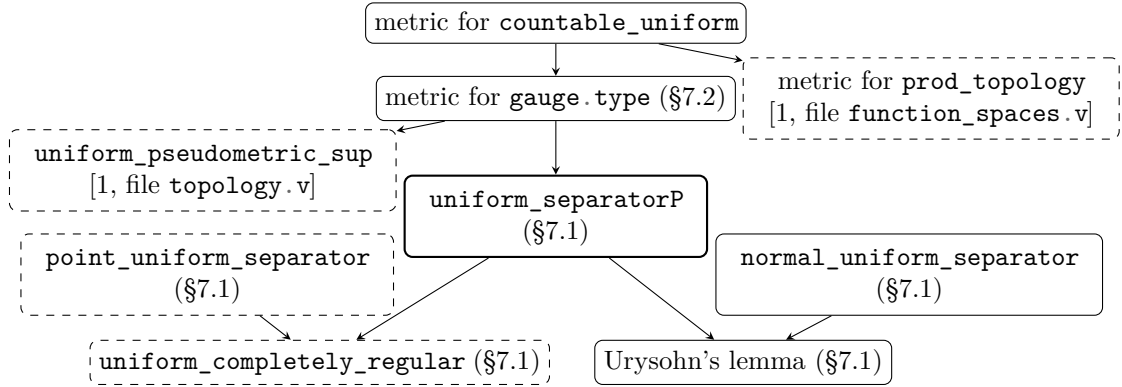
\begin{figure}[t]
\centering
\begin{tikzpicture}
\begin{scope}[every node/.style={draw,fill=white,rounded corners=0.3em},
              every path/.style={-stealth},
              boundedgen/.style={dashed,blue}]
  \node (urysohn) {Urysohn's lemma (\sec{\ref{sec:uniformseparator}})};
  \node (uniform_completely_regular)[left of=urysohn,xshift=-13em,dashed] {\coqin{uniform_completely_regular} (\sec{\ref{sec:uniformseparator}})};
  \node (uniform_separatorP) [above left of=urysohn, yshift=3.5em, xshift=-5em,thick] {\begin{tabular}{c}\coqin{uniform_separatorP}\\ (\sec{\ref{sec:uniformseparator}}) \end{tabular}};
  \node (gauge_metrics) [above of=uniform_separatorP,yshift=1.6em] {metric for \coqin{gauge.type} (\sec{\ref{sec:uniformsep}})};
  \node (countable_uniform) [above of=gauge_metrics] {metric for \coqin{countable_uniform}};
  \node (countable_product_metrizable) [right of=gauge_metrics,xshift=11.2em,dashed] {\begin{tabular}{c}metric for \coqin{prod_topology}\\ \cite[file \coqin{function_spaces.v}]{analysis}\end{tabular}};
  \node (point_uniform_separator) [above left of=uniform_completely_regular,yshift=1.3em,xshift=-2.2em,dashed] {\begin{tabular}{c}\coqin{point_uniform_separator}\\ (\sec{\ref{sec:uniformseparator}}) \end{tabular}};
  \node (normal_uniform_separator) [above right of=urysohn,yshift=1.3em,xshift=4.2em] {\begin{tabular}{c}\coqin{normal_uniform_separator}\\ (\sec{\ref{sec:uniformseparator}}) \end{tabular}};
  \node (uniform_metric_sup) [below left of=gauge_metrics,xshift=-11.2em,yshift=-0.7em,dashed] {\begin{tabular}{c}\coqin{uniform_pseudometric_sup}\\ \cite[file \coqin{topology.v}]{analysis}\end{tabular}};
  \draw (normal_uniform_separator) -- (urysohn);
  \draw (uniform_separatorP) -- (urysohn);
  \draw (uniform_separatorP) -- (uniform_completely_regular);
  \draw (point_uniform_separator) -- (uniform_completely_regular);
  \draw (gauge_metrics) -- (uniform_separatorP);
  \draw (countable_uniform) -- (gauge_metrics);
  \draw (countable_uniform) -- (countable_product_metrizable);
  \draw (gauge_metrics) -- (uniform_metric_sup);
\end{scope}
\end{tikzpicture}
\caption{Overview of a novel proof of Urysohn's lemma and derived results\\
\footnotesize{(dashed nodes are derived results that are not relevant to the \LDT{})}}
\label{fig:ury_graph}
\end{figure}

\subsection{Existence of uniform separators}
\label{sec:uniformsep}

The proof of \coqin{uniform_separatorP} from the previous section
(\sect{\ref{sec:uniformseparator}}) follows other intermediate lemmas.
If $T$ is a uniform space with a countable basis for its uniformity
(i.e., the set of entourages of $T$ is the upward closure of a
countable subset of entourages, see \coqin{countable_uniformity}
\cite[file \coqin{topology.v}]{analysis}), then $T$ has a
pseudometric~\cite{lesseig1963}.
The construction of the metric is rather involved, but is only done
once.  Then, whenever we want a function into the reals, we construct
a suitable uniformity instead, and rely on this result to guarantee
such a function.
This has several useful consequences, such as proving countable
products of metric spaces are metrizable (Fig.~\ref{fig:ury_graph})
More importantly, the metrizability of uniform spaces lets us build
the so-called \newterm{gauge metric}.
Given an entourage $E$, we get a metric for the uniform structure
generated by $E_0=E\cap E^{-1}$,
$E_1 = {\hbox{\coqin{split_ent}}}(E_0)\cap{\hbox{\coqin{split_ent}}}(E_0)^{-1}$, etc.\
(where \coqin{split_ent} was explained in \sect{\ref{sec:preliminaries}}).

The direct part of the proof of \coqin{uniform_separatorP} starts by
building the gauge metric of the separator of~$A$ and~$B$ generated
from~$E$.
Since the initial entourage separates~$A$ and~$B$, we know that there is an
$\varepsilon > 0$ such that for all $x\in A$ and $y\in B$, $y\notin \ball{\varepsilon}{x}$.
We define the extended real-valued function
$d(x,y) \mydef \inf\{ r > 0 \,|\, y\in \ball{r}{x}\}$ (\coqin{edist} in
\cite[file \coqin{normedtype.v}]{analysis}) and the extended real-valued function
$d'(A, z) \mydef \inf\{d(z,a) \,|\, a\in A\}$ (\coqin{edist_inf} in
\cite[file \coqin{normedtype.v}]{analysis}).
The function $d$ is continuous on the gauge uniform space.
Since it is coarser than the topology on $T$, $d$ is continuous on $T\times T$,
and thus $d'$ is continuous on $T$.
It follows that $f(x) \mydef \min(d'(A, x), \varepsilon)/\varepsilon$
is continuous.
It also takes $0$ on $A$, and $1$ on $B$, and has range $[0,1]$, which
means that we have the separating function and thus a proof of
\coqin{uniform_separatorP}.  See lemma \coqin{urysohn_separation} in
\cite[file \coqin{normedtype.v}]{analysis} for details.

We now need to show \coqin{normal_uniform_separator}. 
For that purpose, we need to build a suitable uniform structure on a normal space. 
The uniformity we construct has the following basis:
\begin{minted}{ssr}
Let apxU (UV : set T * set T) : set (T * T) :=
  (UV.2 8\mprod8 UV.2) 8\munion8 ((closure UV.1)8\mcplt8 8\mprod8 (closure UV.1)8\mcplt8).
Let nested (UV : set T * set T) :=
  [/\ open UV.1, open UV.2, A 8\msub8 UV.1 & closure UV.1 8\msub8 UV.2].
Let ury_base := [set apxU UV | UV in nested].
\end{minted}
Most of the work is to show that this is a basis for a uniformity, and
that it is coarser than the topology on $T$.
Then given an $A$ and a $B$ which are closed, disjoint, and non-empty, normality
guarantees an open set $U \supseteq A$ with $U \cap B = \emptyset$. 
Then \mintinline{ssr}{apxU (U, B8\mcplt8)} is a separator for $A$ and $B$ and we have 
our \coqin{uniform_separator}. See \cite[file \coqin{normedtype.v}]{analysis} for details.

\section{Lusin and Tietze theorems and continuous functions are dense in $L^1$}
\label{sec:density}

As sketched in \sect{\ref{sec:ldt_proof}}, one important step to prove
the \LDT{} is to establish that continuous functions are dense in~$L^1$.
As we explained in \sect{\ref{sec:ldt_proof}}, we can get along with a
formal proof considering only functions defined over a bounded set.
As an intermediate step, we formalize Lusin's theorem and Tietze's extension theorem.

\subsection{Lusin's theorem}
\label{thm:lusin}

We place ourselves in a context
with \coqin{R : realType} to represent a type of reals and
where \coqin{mu} is the Lebesgue measure.
Lusin's theorem states that, given a measurable function $f$
over $A$ (a measurable bounded set) and $\varepsilon>0$, there
exists a compact $K\subseteq A$ such that
$\mu(K \setminus A) < \varepsilon$ and $f$ is continuous within $K$
\cite[file \coqin{lebesgue_integral.v}]{analysis}:
\begin{minted}{ssr}
Lemma measurable_almost_continuous (f : R -> R) (A : set R) (eps : R) :
  measurable A -> mu A < +oo -> measurable_fun A f -> 
  0 < eps -> exists K,
    [/\ compact K, K 8\msub8 A, mu (A 8\mdiff8 K) < eps%:E & {within K, continuous f}].
\end{minted}
The notation \coqin{{within _, continuous _}} was explained along
the formalization of subspace topologies in \sect{\ref{sec:subspace}}.
The proof also uses the following lemma that pertains to
subspace topologies as well: if \coqin{f} and \coqin{g} are equal on \coqin{A}
and \coqin{f} is continuous then so is~\coqin{g}. Put formally:
\begin{minted}{ssr}
Lemma subspace_eq_continuous {S : topologicalType} (f g : subspace A -> S) :
  {in A, f =1 g} -> continuous f -> continuous g.
\end{minted}
The proof connects to results presented earlier in this paper:
Egorov's theorem (\sect{\ref{sec:egorov}}) and the (bounded version of
the) inner regularity of Lebesgue measurable
(\sect{\ref{sec:inner_regularity}}).

\subsection{Tietze's extension theorem}
\label{sec:tietze}

Tietze's extension theorem states that in a normal topological space
(normality being already defined in \sect{\ref{sec:urysohn}}), a bounded, continuous,
real-valued function on a closed set can be extended to a bounded, continuous
function on the whole set.
Although we do formalize Tietze's theorem for normal spaces, it should
be noted that the normality condition is incidental to
the main results of this paper; what is relevant here is that the
reals are normal.
Here follows the statement of Tietze's theorem in \analysis{} \cite[file \coqin{numfun.v}]{analysis}:
\label{thm:tietze}
\begin{minted}{ssr}
Context {X : topologicalType} {R : realType}.
Hypothesis normalX : normal_space X.
Lemma continuous_bounded_extension (f : X -> R^o) (A : set X) M :
  closed A -> {within A, continuous f} -> 
  0 < M -> (forall x, A x -> `|f x| <= M) ->
  exists g, [/\ {in A, f =1 g}, continuous g & forall x, `|g x| <= M].  
\end{minted}
The notation \coqin{^o} is only to help type-checking.  Besides
Urysohn's lemma, this proof uses the fact that uniform convergence
preserves continuity (lemma \coqin{uniform_limit_continuous} in
\cite[file \coqin{function_spaces.v}]{analysis}).
The hypothesis \coqin{{within A, continuous f}} is a typical detail
that does not appear in a textbook where this theorem would be
assuming that the function \coqin{f} is continuous on~\coqin{A}.

\subsection{Continuous functions are dense in $L^1$}
\label{sec:cont_dense}

Finally, we arrive at the true goal of this section: the fact that continuous
functions are dense in $L^1$, i.e., that given a function $f$ integrable over
a measurable, bounded set $A$, there exists a sequence of continuous functions
$g_k$, integrable over $A$, such that $|| f - g_k ||_1$ tends towards~$0$:
\label{thm:dense}
\begin{minted}{ssr}
Lemma approximation_continuous_integrable (A : set _) (f : R -> R):
  measurable A -> mu A < +oo -> mu.-integrable A (EFin \o f) ->
  exists g_ : (R -> R)^nat,
    [/\ forall n, continuous (g_ n),
        forall n, mu.-integrable A (EFin \o g_ n) &
        \int[mu]_(z in A) `|(f z - g_ n z)%:E| @[n --> \oo] --> 0].
\end{minted}
The proof uses Tietze's and Lusin's theorems, see \cite[file
\coqin{lebesgue_integral.v}]{analysis}.
As we explained in \sect{\ref{sec:ldt_proof}}, we use the above lemma
to produce a sequence of continuous functions~$g_i$ to be used in the
``bounded version'' of the \LDT{} for a real-valued function restricted
to some ball~$B_k$.
The desired sequence of $g_i$'s is obtained by the above lemma modulo
the technical detail that we need to restrict them to $B_k$ for them to
connect correctly to other lemmas used in the proof of the \LDT{}.

\section{Covering lemmas and the \HL{} maximal inequality}
\label{sec:vitali_hardylittlewood}

As we explained in \sect{\ref{sec:ldt_proof}}, the second important
step to prove the \LDT{} is the \HL{} maximal inequality, i.e., the
fact that, for all locally integrable functions $f$,
$\mu(\{x\,|\,\HLmax{f}(x)>c\})\leq\frac{3}{c}||f||_1$
for all $c>0$.
Its proof relies on a covering lemma typical of measure theory.

\subsection{Vitali's covering lemma}
\label{sec:vitali}

In its finite version, the Vitali covering lemma can be stated as
follows: given a finite collection of balls $B_i$ with $i\in s$, there
exists a subcollection $B_j$ with $j\in D$ of pairwise disjoint balls
such that $\bigcup_{i\in s} B_i \subseteq \bigcup_{j\in D} 3B_j$.
To formalize this statement without committing to a concrete representation for
collection of balls, we represent them by a
function \coqin{B : I -> set R} such that each set satisfies a
predicate \coqin{is_ball}, instead of representing them, say,
as a function returning pairs of a center and a radius.
The approach using the \coqin{is_ball} predicate gives rise to two functions
\coqin{cpoint} and \coqin{radius} returning respectively a center point and
a non-negative radius, when the set is indeed a ball.
The finiteness of the collections is captured by using lists (respectively
\coqin{s} and \coqin{D} below).
\begin{minted}{ssr}
Context {I : eqType}.
Variable (B : I -> set R).
Hypothesis is_ballB : forall i, is_ball (B i).
Hypothesis B_set0 : forall i, B i !=set0.

Lemma vitali_lemma_finite (s : seq I) : { D : seq I | [/\ uniq D,
  {subset D <= s}, trivIset [set` D] B &
  forall i, i \in s -> exists j, [/\ j \in D, B i 8\minter8 B j !=set0,
    radius (B j) >= radius (B i) & B i 8\msub8 3 *` (B j)] ] }.
\end{minted}
The notation \coqin{*`} represents scaling of the radius of a ball,
i.e., \coqin{k *` B} is the open ball with center \coqin{cpoint B} and
radius \coqin{k * radius B}.

We also formalized the infinite version of Vitali's covering lemma
\cite[file \coqin{normedtype.v}]{analysis} and Vitali's theorem
\cite[file \coqin{lebesgue_measure.v}]{analysis},
which are much more involved.  We did not need them to prove
the \LDT{} but they served as a test-bed for using the \coqin{is_ball}
predicate and are anyway often mentioned in connection with
proofs of the \FTC{}.

\subsection{Hardy-Littlewood maximal inequality}
\label{sec:hardylittlewood}

The \HL{} operator is a function that transforms a real-valued
function $f$ into the function
$$\HLmax{f}(x)\mydef\sup_{r>0}\frac{1}{\mu(\ball{r}{x})}\inte{\mu}{y\in
  \ball{r}{x}}{|f(y)|}.$$ Its formal definition uses elements similar to the
ones used when defining Lebesgue points in \sect{\ref{sec:ldt_statement}}:
\begin{minted}{ssr}
Definition HL_max (f : R -> R) (x : R^o) (r : R) : \bar R :=
  (fine (mu (ball x r)))^-1%:E * \int[mu]_(y in ball x r) `|(f y)%:E|.
Definition HL_maximal (f : R -> R) (x : R^o) : \bar R :=
  ereal_sup [set HL_max f x r | r in `]0, +oo[ ].
\end{minted}
The statement of the \HL{} maximal inequality that we explained
informally at the very beginning of this section
(\sect{\ref{sec:vitali_hardylittlewood}}) then translates directly:
\begin{minted}{ssr}
Lemma maximal_inequality (f : R -> R) c :
  locally_integrable [set: R] f -> 0 < c ->
  mu [set x | HL_maximal f x > c%:E] <= (3%:R / c)%:E * norm1 [set: R] f.
\end{minted}
The $L^1$ norm is formalized in the obvious way as the identifier
\coqin{norm1}.  The proof relies on inner regularity
(\sect{\ref{sec:inner_regularity}}) and Vitali's covering lemma
(\sect{\ref{sec:vitali}}).  To establish that the \HL{} operator is
measurable, we also need to develop a theory of lower semicontinuity,
which has been added to \analysis{} on this occasion, see \cite{analysis}
for details.

\section{Applications of the \LDT{}}
\label{sec:applications}

In the previous sections, we have explained the main lemmas (mainly:
continuous functions are dense in $L^1$ and the \HL{} maximal
inequality) used to prove the \LDT{} that we sketched in
\sect{\ref{sec:ldt_proof}}. We are now ready to proceed to direct
applications.

\subsection{The first \lebFTC{}}
\label{sec:ftc}

Recall from \sect{\ref{sec:introduction}} the informal statement of
the first \lebFTC{}: for $f\in L^1$,
$F(x)\mydef\inte{\mu}{t\in]-\infty,x]}{f(t)}$ is differentiable and
$F'(x)\myae f(x)$.
This can furthermore be generalized to intervals of the form $]a, x]$ and $[a, x]$
and stated as a single theorem as follows \cite[file \coqin{ftc.v}]{analysis}:
\begin{minted}{ssr}
Lemma FTC1_lebesgue_pt f a : mu.-integrable [set: R] (EFin \o f) ->
  let F x := (\int[mu]_(t in [set` Interval a (BRight x)]) (f t))%R in
  forall x, a < BRight x -> lebesgue_pt f x ->
  derivable F x 1 /\ F^`() x = f x.
\end{minted}
The variable \coqin{a} has the generic type of an ``interval bound''
and \coqin{BRight} stands for closed bounds on the right
\cite[file \coqin{interval.v}]{mathcomp}.
The predicate \coqin{derivable} is for derivability (\coqin{1} is the
direction) and the notation \coqin{^`()} is for derivatives with
domain~$\mathbb{R}$~\cite[\sec{4.5}]{cohen2018jfr}.
The theorem above connects these notions with the Lebesgue
integral developed in \analysis{} independently~\cite[\sec{6.4}]{affeldt2023jar}.
The proof is standard in that it goes through a generalization of the
\LDT{} where balls are replaced with \newterm{nicely shrinking}
sets~\cite[\sec{II.4.1}]{li2022}.
The statement of the first \FTC{} from \sect{\ref{sec:introduction}} is an
immediate corollary of the above lemma:
\begin{minted}{ssr}
Corollary FTC1Ny f : mu.-integrable setT (EFin \o f) ->
  let F x := (\int[mu]_(t in [set` `]-oo, x]]) (f t))%R in
  {ae mu, forall x, derivable F x 1 /\ F^`() x = f x}.
\end{minted}

\subsection{Lebesgue's density theorem}
\label{sec:lebesgue_density}

Lebesgue's density theorem is another direct consequence of the
\LDT{}. The \newterm{density} of a point $x$ w.r.t.\ a set $A$ is
defined by
$\displaystyle \lim_{r\to 0^+}\frac{\mu(A\cap
  \ball{r}{x})}{\mu(\ball{r}{x})}$. Lebesgue's density theorem states that
almost everywhere the density is 0 or 1:
\begin{minted}{ssr}
Lemma density (A : set R) : measurable A ->
  {ae mu, forall x, mu (A 8\minter8 ball x r) * (fine (mu (ball x r)))^-1%:E
    @[r --> 0^'+] --> (\1_A x)%:E}.  
\end{minted}

\section{Related work}
\label{sec:related_work}

We have been using various documents to formalize the \LDT{}. In particular,
the main lines are drawn from lecture notes by Bowen~\cite{bowen2014}.  For the
proofs of the \HL{} maximal inequality and the proof of the \LDT{},
we used books by Li~\cite{li2022} and
Schwartz~\cite{schwartz1997}. Surely, the same contents can be found
elsewhere.

Several lemmas that we discussed can also be found in Mathlib~\cite{mathlib2020cpp}.
Of course, the proof of Urysohn's lemma in Mathlib \cite{mathliburysohn} is
different from ours, which is original, as we explained in
\sect{\ref{sec:urysohn}}.
Tietze's extension theorem in Mathlib \cite[class \coqin{TietzeExtension}]{mathlibtietze}
has a similar statement and a similar proof.
The statement of the \LDT{} in Mathlib \cite{mathlibdifferentiation}
is more general than ours: it allows the domain of the function to be
an arbitrary metric space, the measure can be any locally finite
measure, the codomain can be any normed abelian group, and the balls
(used in \coqin{davg} in \sect{\ref{sec:ldt_statement}}) can be
replaced by an arbitrary Vitali family.
The \LDT{} in Mathlib is also used to prove a generic version of Lebesgue's
density theorem~\cite{mathlibdensity}\cite[\sec{3.2}]{nash2023itp}.

\def\myapprox{$\approx\,$}

\begin{table}
\begin{tabular}{lllll}
\hline
{\it Supporting theories} & Section & l.o.c. & & file in~\cite{analysis} \\
\hline
Subspaces & \sect{\ref{sec:subspace}} & N.A. & & \coqin{topology.v} \\
Uniform convergence & \sect{\ref{sec:uniform_cvg}} & N.A. & & \coqin{function_spaces.v} \\
\hline
\hline
{\it Main lemmas} & Section & l.o.c. & & file in~\cite{analysis} \\
\hline
Egorov's thm & \sect{\ref{sec:egorov}} & \myapprox 87 & (2 lemmas) & \coqin{lebesgue_measure.v} \\
Outer regularity & \sect{\ref{sec:outer_regularity}} & \myapprox 61 & (1 lemma) & \coqin{lebesgue_measure.v} \\
Inner regularity & \sect{\ref{sec:inner_regularity}} & \myapprox 118 & (4 lemmas) & \coqin{lebesgue_measure.v} \\
Lusin's thm & \sect{\ref{thm:lusin}} & \myapprox 108 & (3 lemmas) & \coqin{lebesgue_integral.v} \\
Tietze's extension thm & \sect{\ref{sec:tietze}} & \myapprox 108 & (3 lemmas) & \coqin{numfun.v} \\
Density of cont.\ functions & \sect{\ref{sec:cont_dense}} & \myapprox 118 & (3 lemmas) & \coqin{lebesgue_integral.v} \\
Finite Vitali's covering lem. & \sect{\ref{sec:vitali}} & \myapprox 75 & (2 lemmas) & \coqin{normedtype.v} \\
Hardy-Littlewood max.\ ineq.\ & \sect{\ref{sec:hardylittlewood}} & \myapprox 180 & (6 lemmas) & \coqin{lebesgue_integral.v} \\
Urysohn's lemma & \sect{\ref{sec:uniformseparator}} & \myapprox 165 & (11 lemmas) & \coqin{normedtype.v} \\
Lebesgue Differentiation thm & \sect{\ref{sec:ldt_proof}} & \myapprox 143 & (3 lemmas) & \coqin{lebesgue_integral.v} \\
First FTC & \sect{\ref{sec:ftc}} & \myapprox 265 & (4 lemmas) & \coqin{ftc.v} \\
Lebesgue's density thm & \sect{\ref{sec:lebesgue_density}} & \myapprox 69 & (1 lemma) &  \coqin{lebesgue_integral.v} \\
\hline
\multicolumn{2}{r}{Total ({\it Main lemmas})} & \myapprox 1,497 \\
\end{tabular}
\caption{Estimated lines of code for the formalization of the \LDT{} and its direct applications\\
  {\footnotesize (the column l.o.c.\ contains the number of lines of
    code in proof scripts for the main proof and intermediate lemmas
    (including statements); these numbers are
    approximations because, among other reasons, where one draws the
    line between intermediate lemmas and the supporting theories can
    be arbitrary) } }
\label{tab:dev_overview}
\end{table}

The \FTC{} has already been the target of several formalizations in
proof assistants.
It can be found in \coq{} but for the Riemann integral in a
constructive setting \cite[\sec{6}]{cruzfilipe2002types}.
NASAlib does not feature the first \lebFTC{} but an elementary version
(for a $C^1$ function) of the second \FTC{} \cite[file
\coqin{lebesgue_fundamental.pvs}]{pvslib}, which can actually be
obtained from the first \lebFTC{} as a corollary.
Isabelle/HOL features the first \lebFTC{} but for continuous functions
whereas we prove it for integrable
functions~\cite[\sec{3.7}]{avigad2017jar}.
Mathlib features several variants of the first \FTC{}; many require
integrability and continuity at the endpoints but establish strict
differentiability~\cite{mathlibfunthmcalculus}.
%
%
They stem from a lemma analogous to a strengthening of the \LDT{} with
nicely shrinking sets~\cite{mathlibsetintegral}.
In other words, we are able to match our lemmas with Mathlib lemmas
but statements and proofs are organized in a different way.
However, it must be said that Mathlib's statements are admittedly more
general than ours in many respects.
One reason is that \analysis{} has started to use \hb{} pervasively
only recently. Before that, mathematical structures were manually
encoded with \newterm{packed classes}~\cite{garillot2009tphols}: this
was making modifications very difficult in practice.
With \hb{}, we believe that introduction of, say, Banach spaces,
should be a matter of engineering because most of our proofs are
textbook, because we do not abuse the fact that we are working on the
real line, and because our development is short enough to be
refactored (see Table~\ref{tab:dev_overview} for a concrete size
estimation).

\section{Conclusions}
\label{sec:conclusions}

In this paper, we provided a comprehensive overview of the \LDT{}.  We
started with a formal statement and a proof overview (in
\sect{\ref{sec:ldt_statement}} and in \sect{\ref{sec:ldt_proof}}) to
plan the whole development (as summarized in Fig.~\ref{fig:ldt_graph}).
Before being able to even state the first intermediate lemmas, we
needed to extend \analysis{} with in particular new topological
constructs in \sect{\ref{sec:extension_analysis}}.  This made it
possible to formalize the needed basic lemmas from measure theory in
\sect{\ref{sec:more_measure_theory}}.  Among the needed lemma, we
formalized in particular a novel proof of Urysohn's lemma in
\sect{\ref{sec:urysohn}}.
We used all this material to formalize the main steps of the proof of the
\LDT{}: the density of continuous functions in $L^1$ in
\sect{\ref{sec:density}} and \HL{} maximal inequality in
\sect{\ref{sec:vitali_hardylittlewood}}.
Our formalization of the \LDT{} was completed by two applications in
\sect{\ref{sec:applications}}, which include the first proof of the
first \lebFTC{} for the \coq{} proof assistant.

In the end, we provide in a single document a complete overview of an
important theorem.
We believe that this experiment also concretely illustrates an
important aspect of formalization of mathematics: the \LDT{}, like the
uniform separators of \sect{\ref{sec:uniformseparator}}, are examples
of results whose formalization should be prioritized because, though
technical, they are generic intermediate results from which important
results can be obtained as corollaries (here, the first \FTC{} and
Urysohn's lemma).
More pragmatically, we hope that this overview also contributes to
documenting formalization of real analysis with \analysis{}, for
example by explaining the use of \hb{} to develop topology.  We think
that we demonstrated that \analysis{} is already a rich library and
also a useful tool to formalize mathematics. As a matter of fact, we
could use it to revisit Urysohn's lemma by producing an original
proof.

As for future work, we are now working on the second \lebFTC{} whose
most general form deals with \newterm{absolutely continuous
  functions}~\cite{mohanarangan2021}, using as the main ingredient the
Radon-Nikod\'ym theorem already available in
\analysis{}~\cite{ishiguro2024cs} and the recently developed theory of
bounded and total variation \cite[file \coqin{realfun.v}]{analysis}.

\paragraph{Acknowledgements}
The authors are grateful to A.\ Bruni, C.\ Cohen, Y.\
Ishiguro, and T.\ Saikawa for their inputs.
This work has benefited from feedback gained during the \analysis{}
development meetings.
The authors would like to thank the anonymous referees of ITP 2024 for
their careful and informative review, as well as the program committee
of JFLA 2024 where a preliminary version of this work was
presented~\cite{affeldt2024jfla}.
The first author acknowledges support of the JSPS KAKENHI Grant
Number 22H00520.

\bibliographystyle{abbrv}
\bibliography{lebdiffthm.bib}

\end{document}